\documentclass[%
 aip,
 amsmath,amssymb,
 reprint,%
]{revtex4-1}

\usepackage{graphicx}
\usepackage{dcolumn}
\usepackage{bm}

\usepackage[utf8]{inputenc}
\usepackage[T1]{fontenc}
\usepackage{mathptmx}
\usepackage{etoolbox}

\makeatletter
\def\@email#1#2{%
 \endgroup
 \patchcmd{\titleblock@produce}
  {\frontmatter@RRAPformat}
  {\frontmatter@RRAPformat{\produce@RRAP{*#1\href{mailto:#2}{#2}}}\frontmatter@RRAPformat}
  {}{}
}%
\makeatother
\begin{document}

\preprint{AIP/123-QED}

\title{Dynamics of the fractional quantum Hall edge probed by stroboscope measurements of trions}

\author{Akinori Kamiyama}
\affiliation{Department of Physics, Tohoku University, Sendai 980-8578, Japan}
\author{Masahiro Matsuura}
\affiliation{Department of Physics, Tohoku University, Sendai 980-8578, Japan}
\author{John N. Moore}
\affiliation{Department of Physics, Tohoku University, Sendai 980-8578, Japan}
\author{Takaaki Mano}
\affiliation{National Institute for Materials Science, Tsukuba, Ibaraki 305-0047, Japan}
\author{Naokazu Shibata}
\affiliation{Department of Physics, Tohoku University, Sendai 980-8578, Japan}
\author{Go Yusa}
\email{yusa@tohoku.ac.jp}
\affiliation{Department of Physics, Tohoku University, Sendai 980-8578, Japan}

\date{\today}

\begin{abstract}
     By using observations from pump-probe stroboscopic confocal microscopy and spectroscopy,
we demonstrate the dynamics of trions and the fractional quantum Hall edge on the order of $\sim1$~ps. 
The propagation of the quantum Hall edge state excited by a voltage pulse is detected as a temporal change in reflectance in the downstream edge probed by optical pulses synchronized with the voltage pulse. The temporal resolution of such stroboscopic pump-probe measurements is as fast as the duration time of the probe pulse ($\sim1$~ps). This ultra-fast stroboscope measurement enables us to distinguish between the normal mode of edge excitation, known as the edge magneto-plasmon or charge density wave, and other high-energy non-linear excitations. This is the only experimental method available to study the ultra-fast dynamics of quantum Hall edges, and makes it possible to derive the metric tensor $g_{\mu \nu}$ of the $(1+1)=2$-dimensional curved spacetime in quantum universe and black hole analogs implemented in the quantum Hall edge.
\end{abstract}

\maketitle

In general, a topological material comprises a gapped bulk and a conducting edge \cite{thouless82, wen95, hasan10}. 
The quantum Hall (QH) state, which forms when electrons constrained in two dimensions (2D) are exposed to a strong perpendicular magnetic field $B$, was the earliest example of such topological
insulators. When the Landau-level filling factor $\nu = \frac{h}{eB} n_e$, which is the ratio of electron density $n_e$ and the flux quanta density $n_\phi=\frac{eB}{h}$, is an integer or rational fraction, the bulk of the system develops a gap to form integer \cite{klitzing80} and fractional QH (FQH) states \cite{tsui82}. Here $h$ and $e$ stand for the Planck constant and elementary charge, respectively. 
The excitation of the edge, also known as the edge magneto-plasmons (EMPs) or charge density wave \cite{wassermeier90, aleiner94}, has chirality and propagates unidirectionally\cite{ashooriPRB92, ernst97, kamata14, hashisaka17} because $B$ breaks the time-reversal symmetry. The excitation of the edge has sparked new interest due to both its unique
prospective applications and fundamental physics \cite{bid10, sabo17, nakamura20, bartolomei20}. Topics include flying qubit quantum computation \cite{yamamotoNatNano12, yangPRB16, dlaskaQST17, elman17, shimizuPRB20}, quantum energy teleportation\cite{yusaPRA11, matsuuraAPL18, hottaPRA14}, and quantum field simulators which simulate the quantum universe and black holes in $(1+1)=2$-dimensional spacetime \cite{hottaPRA14, hottaPRD22}. Here, $(1+1)$ represents each spatial and temporal dimension as one.  It is necessary to learn about the dynamics of the QH edge to create a simulator for a toy model of an expanding universe. In cosmological language, learning about the QH edge dynamics corresponds to obtaining a tensor field of the metric $g_{\mu \nu}$ \cite{hottaPRD22} which describes how the spacetime is curved.

\begin{figure*}
    \begin{center}
    \includegraphics[scale=1, bb= -15 645 595 840]{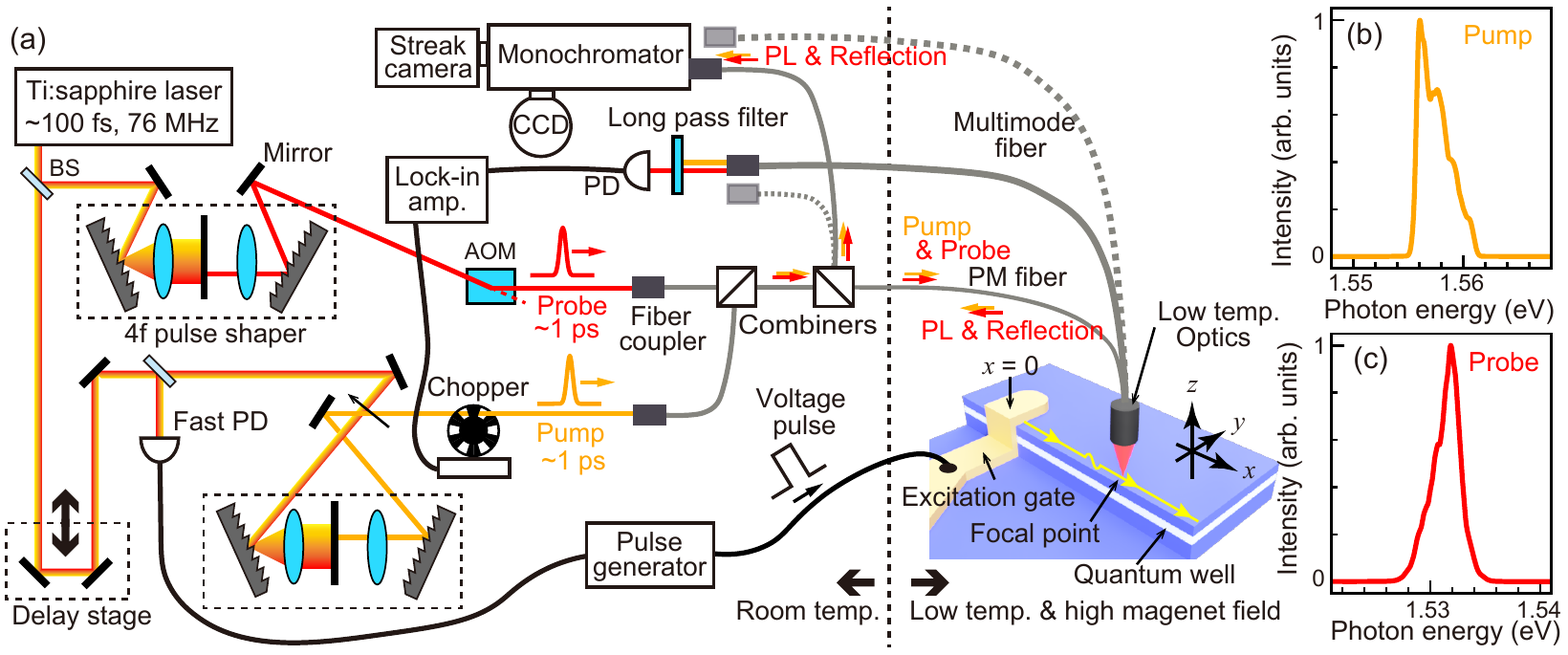}
    \end{center}
\caption{(a) Confocal stroboscopic microscopy and spectroscopy experimental setup diagram. Yellow and orange lines and arrows, respectively, designate the pump and probe pathways. Here PD, AOM, and PM fiber denote photo detector, acoustic optical modulator, and polarization maintaining fiber. Black lines denote coaxial cables. The edge of the mesa is defined as $y=0$, and the center of the excitation gate is specified as $x=0$. The width of the excitation gate along the $x$-axis is $\sim 5$~$\mu$m. (b) and (c) The typical spectra of (b) pump and (c) probe pulses measured by the monochromator and the CCD.}%
\label{fig:fig1}%
\end{figure*}

Scanning stroboscopic photoluminescence (PL) spectroscopy was recently successful in imaging the propagating QH edge in the $\nu=1/3$ FQH regime in real-space and real-time\cite{kamiyamaPRR22}.  The excitation of the edge was probed using the PL intensity and energy of trions, which are the bound states of two electrons and a hole\cite{wojsPRB00, yusaPRL01}. Although the lifetime of trions in the FQH regime has not been reported, it is expected to be comparable to the lifetime of trions in $B$, which is on the order of $\sim 100$~ ps\cite{finkelsteinPRB98}. Because the PL signal for the stroboscopic measurement is averaged across time the time resolution of the stroboscopic PL was anticipated to be limited to the order of $\sim 100$~ps\cite{kamiyamaPRR22}. In this paper, by analyzing the lifetimes of trions in the QH regime using time-resolved PL measurement and pump-probe reflection measurements, we present a method to enhance the time resolution and show excitation of the edge different from the Tomonaga-Luttinger liquid type of excitation\cite{yoshiokaQH}.

The experimental setup consists of room and low-temperature regions (Fig. $1$). A $\sim 100$-fs mode-locked Ti:sapphire laser with a repetition rate of $76$~MHz splits into the pump and probe paths. To reduce wavelength dispersion in polarization-maintaining (PM) fibers, we stretch the pulse duration to $\sim1$~ps using a 4f pulse shaper\cite{weiner00, anghelPRB16} consisting of a slit and two sets of $1200$~/mm-gratings and lenses. The typical spectra of pump and probe lasers have $\sim 3$-meV full width at half maximum (FWHM) (Figs. 1(b) and 1(c)). 
A part of the pump pulse is introduced to a fast photo detector (PD) in the room-temperature region to synchronize with the pulse generator used to excite the QH edge by a voltage pulse. 
The pump pulse is introduced into a PM fiber combined with another PM fiber introduced to the low-temperature region by a combiner. For pump-probe measurement, the pump pulse is chopped by the optical chopper.
The probe pulse is passed through an acoustic optical modulator (AOM) to control the power of the laser. 
By adjusting the 4f pulse shaper's slit, the peak photon energy of the probe pulse is tuned such that it resonates with the tron's PL peak energy. 
The probe pulse is introduced to a PM fiber and illuminates the sample through the low-temperature optics. 
Both pump and probe travel through the same PM fiber and have the same focal position on the sample without noticeable extension of the pulse duration, enabling us to attain high space and time resolution.

\begin{figure}[b]
    \begin{center}
    \includegraphics[scale=1.0, bb = 0 662 595 840, clip]{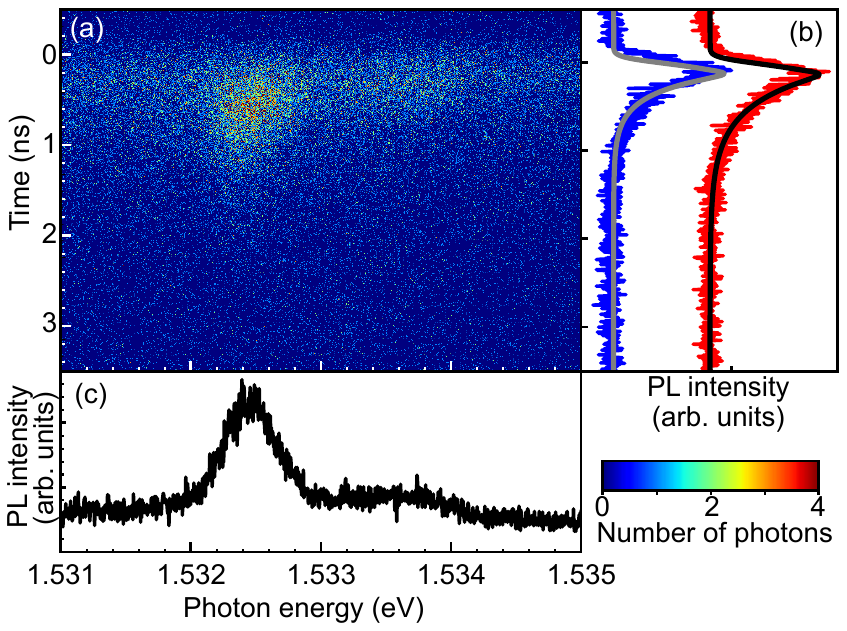}
    \end{center}
    \caption{(a) Microscopic PL spectra obtained in the QH bulk region of $\nu = 2/3$ at $V_\mathrm{b}=-0.35$~V plotted as a function of time at $B=9$~T and a temperature $T\sim 50$~mK. The streak camera uses a color-coded system to display the photon counts. (b) Time-resolved PL intensity for singlet (red) integrated over energy between $1532.326$ and $1532.580$~eV and for triplet (blue) integrated between $1533.361$ and $1533.842$~eV (offset for clarity). Fitting functions for 
    $I f(t-t_0,\tau_\mathrm{d})$
    are represented by the black and gray lines. (c) PL spectra that have been integrated over a time range of $-0.12$ to $0.87$~ns.
}%
    \label{fig:fig2}%
    \end{figure}

The same PM fiber, or a multi-mode fiber, are used to gather both the PL and reflected light. The PM fiber is split by another combiner, and one of the PM fibers can be selectively connected to one of the two detection systems. The first system is for the stroboscopic PL measurement\cite{kamiyamaPRR22} and for the time- and energy-resolved measurements using the streak camera. The system comprises a monochromator outfitted with a streak camera and charge-coupled device (CCD) detector. The time resolution of the streak camera itself is $\sim20$~ps, but that of the total system can be larger than $20$~ps because the optical path in the monochromator is not optimized \cite{schiller80}. A PD connected to a lock-in amplifier makes up the second system. This system is used for stroboscopic pump-probe reflection measurement, which we will explain in detail later. By switching the coupling of the PM and multi-mode fibers at room temperature, the streak camera, stroboscopic PL, and pump-probe reflection measurements can be chosen (see gray dotted lines in Fig.~$1$(a)).

The wafer used for the experiment is a $15$-nm GaAs/AlGaAs quantum well (QW) (Fig.~$1$(a)) grown on an $n+$ substrate, which functions as a back gate to control $n_e$ by applying a back gate voltage $V_\mathrm{b}$. Chemical etching is used to define the sample edge. A front gate electrode made with titanium and gold functions as an excitation gate, and is connected to the pulse generator. The low-temperature optics consists of fiber collimators, beam splitters, a polarizer, and an objective lens with the numerical aperture of 0.55 \cite{kamiyamaPRR22,hayakawa, moorePRB16, moorePRL17, moorePRB18}. Piezoelectric stages can move the sample in the $x$-, $y$-, and $z$-directions relative to the low-temperature optics in a low temperature ($40$~mK) and high $B(\leq 14$~T) environment without heating the sample. 

We begin by displaying time-resolved microscopic PL spectra (Fig.~$2$). PM fiber is used to transmit light from the pump laser onto the sample in a region away from the sample edge (a QH bulk region). Using the same PM fiber, the streak camera attached to the monochromator with confocal geometry detects PL from this region (Fig.~$1$(a)). $\nu$ is set to be $2/3$. The PL spectrum consists of two peaks originating from singlet and triplet trions\cite{wojsPRB00, yusaPRL01}. 
The rate equation describes the dynamics of the trion population undergoing generation and recombination. 
Although the laser pulse resembles a delta function in time, the response of the illumination and detection parts of the system is broadened.
To extract the lifetime of trions, we calculate the convolution of the population decay function and the impulse response of the optical system.
\begin{subequations}
\begin{align}
f(t,\tau) & = \int_{-\infty} ^\infty
N(t^{\prime},\tau)I_{\mathrm{sys}}(t-t^{\prime})dt^{\prime} \label{eq1a}\\
N(t,\tau) &=\left\{ \begin{array}{cc}
0    &  (t < 0)     \\
 e^{-\frac{t}{\tau}}  & (0 \leq t)    
\end{array} \right.\label{eq1b}\\
I_{\mathrm{sys}} (t)&=\frac{1}{\sqrt{2\pi} t_\mathrm{s}} \exp{ \left(  -\frac{t^2}{2 t_\mathrm{s}^2} \right)}, \label{eq1c}\
\end{align}%
\end{subequations}
where $N(t,\tau)$ is the number of trions with a lifetime $\tau$, and $I_{\mathrm{sys}} (t)$ is the impulse response of the system with a full width at half maximum (FWHM) of $2t_\mathrm{s}\sqrt{2\mathrm{ln}2}$. The $t_\mathrm{s}$ corresponds to the standard deviation.
The following function containing the error function $\mathrm{erf}(t)$ can be used to compute the convolution.
\begin{subequations}
\begin{align}
f(t,\tau) &= \frac{1}{2}\left(\mathrm{erf} \left(\frac{t\tau - t_\mathrm{s}^{2}}{\sqrt{2}t_\mathrm{s} \tau}  \right) + 1  \right) \exp \left( \frac{-2t\tau + t_\mathrm{s}^2}{2\tau^2} \right) \label{eq2a}\\
\mathrm{erf}(t) &= \frac{2}{\sqrt{\pi}}\int_0 ^t e^{-x^2} dx. \label{eq2b}\
\end{align}%
\end{subequations}

The time dependence of the PL 
can be fitted by 
$I f(t-t_0,\tau_\mathrm{d})$, where $I$ is a coefficient and $t_0$ is the time when the pump pulse arrives at the sample. 
$\tau_\mathrm{d}$, the decay time
for the singlet and triplet, is respectively $361$ and $214$~ps (Fig.~2(b)). 
The speed of the edge is experimentally reported as on the order of $10^{4-5}$~m$/$s\cite{kamata14,hashisaka17,matsuuraAPL18,kamiyamaPRR22}. Therefore, the edge excitation propagates $2\sim30$~$\mu$m within $200 \sim 300$~ps. 
However, the time resolution of the stroboscopic PL measurement is limited by the duration of the PL intensity peak which is mostly determined by the 200 to 300 ps decay of the trion population.
This means that the time resolution needed to fully capture the dynamics of edge propagation is insufficient.

\begin{figure}
    \begin{center}
    \includegraphics[scale=1.0, bb = 0 615 595 842, clip]{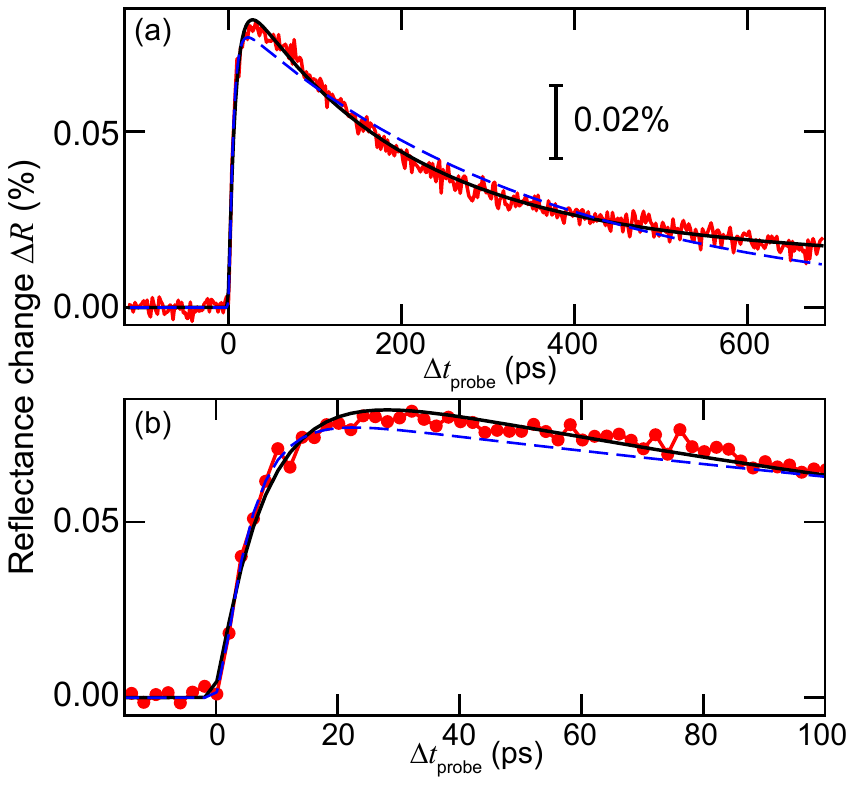}
    \end{center}
    \caption{(a) Reflectance change $\Delta R$  as a function of the time difference $\Delta t_{\mathrm{probe}}$ between the pump and probe pulses (red line) at $\nu=2/3$ ($B=11.5$~T and $T=60$~mK). Their fitting curves (dashed blue line: one-$\tau_\mathrm{d}$ model and black line: two-$\tau_\mathrm{d}$ model) are calculated 
    (see body text).
    The measurement position is $x=20$~$\mu$m. The photon energy of the pump and probe pulse is 1.56--1.57 and $1.53$~eV, and the laser power of the pump and probe pulses is $\sim$200--400 and $\sim100$~nW, respectively. Both the pump and probe pulses have a $\sim 1$~ps duration. The integrated time per data point is $40$~s. (b) Magnified view of Fig.~3(a) around $\Delta t_{\mathrm{probe}}=0$. Measurement data is plotted in red circles.
    }%
    \end{figure}
    
Owning to the Kramers–Kronig relation, reflectance measurement of light which resonates with trions is equivalent to transmittance measurement, and provides the relaxation time of trions. Therefore, we conduct a measurement of the pump-probe reflectance as a probe in place of the PL measurement to enhance the time resolution of the space- and time-resolved measurement. In addition to the pump pulse, we introduce a probe pulse. The probe pulse's peak energy is set so that it will resonate with the trion's PL peak ($1.53$~eV). The intensity of the pump pulse is modulated by a chopper with a frequency of $1$~kHz, and the modulation component of probe pulse intensity reflected from the sample is detected by a lock-in amplifier connected to the PD (Fig.~1(a)). The long pass filter ($<1.54$~eV) in front of the PD blocks the pump light. The time difference $\Delta t_{\mathrm{probe}}$ between the pump and probe pulses can be changed by the delay stage.

 We measure the reflectance change $\Delta R$ as a function of $\Delta t_{\mathrm{probe}}$ at $\nu=2/3$ close to the sample edge ($x=20$, $y=1$~$\mu$m). $\Delta R$ increases rapidly by as much as $~0.08\%$ at $\sim 20$~ps; then it decays gradually (Fig.~$3$). To fit the $\Delta R$ dependence on $\Delta t_\mathrm{probe}$, 
we used one-$\tau_\mathrm{d}$ and two-$\tau_\mathrm{d}$ models which presuppose one and two decay processes for trions, respectively. 
The fitting function for the one-$\tau_\mathrm{d}$ model is $-If(\Delta t_\mathrm{probe}-t_0,\tau_\mathrm{r})+If(\Delta t_\mathrm{probe}-t_0,\tau_\mathrm{d})$ and that for the two-$\tau_\mathrm{d}$ model is $-(I_1+I_2)f(\Delta t_\mathrm{probe}-t_0,\tau_\mathrm{r})+I_1f(\Delta t_\mathrm{probe}-t_0,\tau_\mathrm{d1})+I_2f(\Delta t_\mathrm{probe}-t_0,\tau_\mathrm{d2})$. Here, we consider a rise time $\tau_\mathrm{r}$, which corresponds to the relaxation time from initially high-energy states of photo-excited electrons and holes to the trion state. This is important because the time resolution of our pump-probe reflectance measurement is comparable to $\tau_\mathrm{r} < \sim 10$~ps (Fig.~3(b)).
The optimized parameters for the one-$\tau_\mathrm{d}$ model are $\tau_\mathrm{r}=5.2$~ps,  $\tau_\mathrm{d}=359$~ps, and $I = 0.0828$~\% (dashed blue curves in Fig.~$3$). Compared to the one-$\tau_\mathrm{d}$ model, the two-$\tau_\mathrm{d}$ model with $\tau_\mathrm{r}=8.0$~ps, $\tau_\mathrm{d1}=179$~ps, $\tau_\mathrm{d2}=2018$~ps, $I_1=0.0732$~\%, and $I_2=0.0225$~\% (black curves) clearly fits better than the one-$\tau_\mathrm{d}$ model. Here, $\tau_\mathrm{d1}$ can be attributed to the relaxation time of trions. $\tau_\mathrm{d2}$ may be related to the spin decay time \cite{anghelPRB16}, but further investigation is required for more details. 

Around $\Delta t_{\mathrm{probe}}=0$, there is a measurable change in $\Delta R$ within the $2$-ps time step of our measurement (see Fig.~$3$(b)). 
This fact shows that the pump-probe reflectance measurement's time resolution has been enhanced to the same order as the laser pulse duration time ($\sim 1$~ps).

We integrate the pump-probe reflectance measurement into the stroboscopic measurement, where the electrical excitation of the edge by voltage pulses and subsequent detection by optical light pulses are coordinated. In this stroboscopic measurement, the population of trions is probed within the duration of the probe pulse ($\sim1$~ps). The pulse generator electrically controls $\Delta t$, which is the time difference between the pump and voltage pulses. $\Delta t_{\mathrm{probe}}$ is fixed to $20$~ps such that $\Delta R$ becomes maximum (Fig.~$3$(b)). The voltage pulse is a square pulse with a $150$~mV amplitude and a $0.5$~ns duration (Fig.~$4$(a)). In Fig.~$4$(b) we show $\Delta R$ as a function of $\Delta t$ at several points of varying $x$ and fixed $y$. 

\begin{figure}
    \begin{center}
    \includegraphics[scale=1.0, bb = 0 397 595 841, clip]{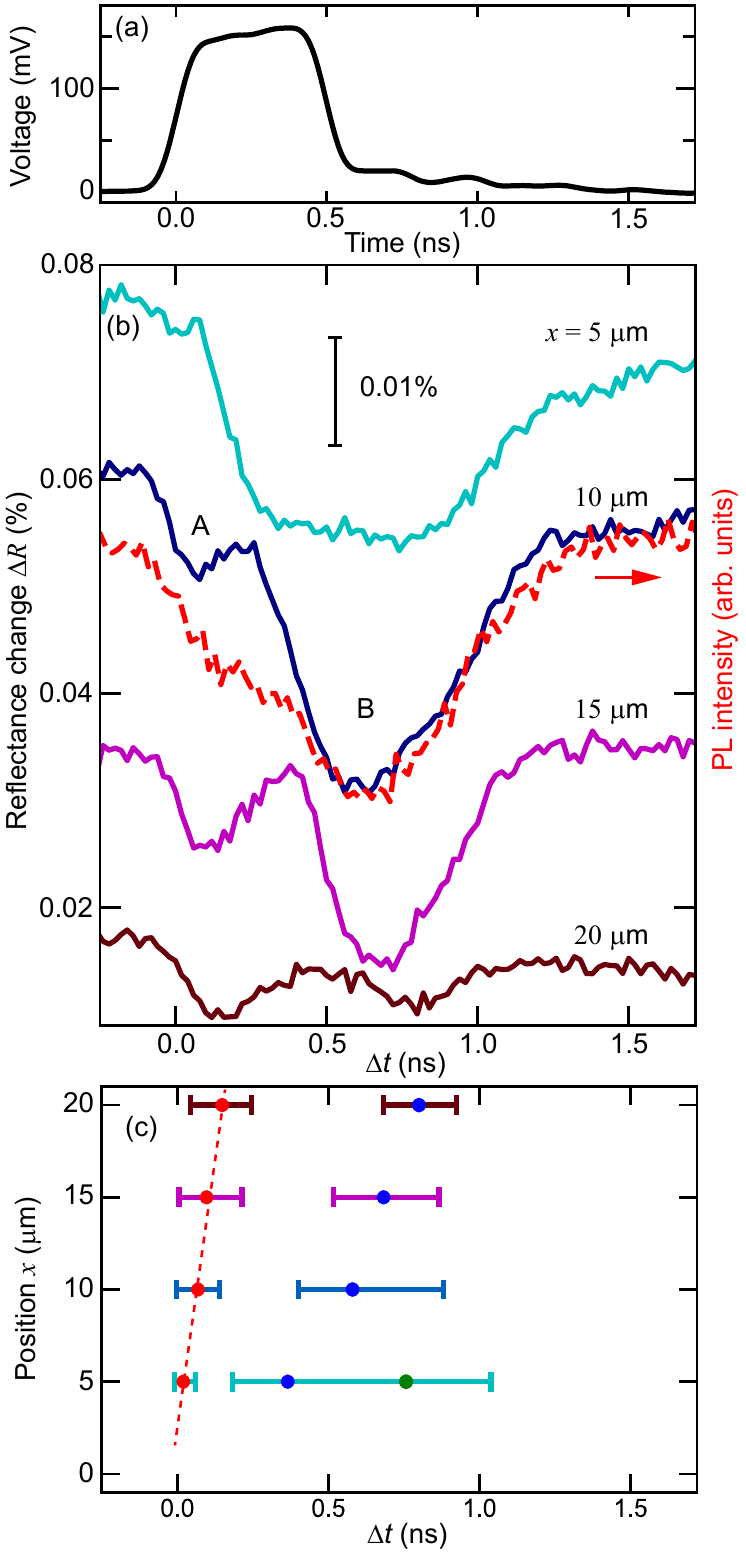}
    \end{center}
    \caption{(a) Voltage waveform applied to the excitation gate to excite the edge. An oscilloscope and pulse generator were connected to measure it. Typical rise and fall time is $\sim 80$~ps for $20-80\%$ (b) Reflectance change $\Delta R$ (solid lines for the left $y$-axis) and PL intensity (red dotted line for the right $y$-axis) as a function of $\Delta t$ (offset for clarity). The PL intensity was measured at $x=10$~$\mu$m, and the $\Delta R$ was measured at $x=5$, $10$, $15$, and $20$~$\mu$m. For all measurements the position of $y$ was $1$~$\mu$m. The integration time is $400$~s per point for $\Delta R$. (c) $x$ as a function of the $\Delta t$ value where dips are observed in the traces of Fig.~$4$(b). The horizontal bars represent $50\%$ changes in $\Delta R$.  
    }%
    \end{figure}

At $x=10$~$\mu$m, small and large dips are observed around $\Delta t=0.1$ and $0.6$~ns labeled as A and B, respectively(Fig.~$4$(b)).
Similar dips are seen in the PL intensity, but the PL dips are wider and more overlapping (dashed red curve in Fig.~$4$(b)). This can be attributed to the difference in the time resolution. In stroboscopic PL measurement, the time resolution is limited by the lifetime of  trions excited by the pump pulse laser. In contrast, the time resolution of the stroboscopic pump-probe reflection measurement is the duration of the probe pulse, and is unrelated to the trion lifetime. Moreover, because the transient of the PL intensity is temporally asymmetric, the broadening of the intensity in the stroboscopic PL measurement is also asymmetric: in Fig. 4(b) at $t<0$~ns, $\Delta R$ is constant  before the voltage pulse is applied, whereas the PL intensity already starts to change.
Therefore, the time resolution of the stroboscopic pump-probe reflection measurement is superior to that of the stroboscopic PL measurement.

The $\Delta t$ at which the dips A and B appear increases as $x$ increases.
In the case of dip A, $\Delta t$ at which dip A appears is proportional to $x$ and the slope of $x$ v.s. $\Delta t$ (red dotted line in Fig.~4(c)) is $\sim 1 \times 10^{5}$~m/s, which is in good agreement with the edge excitation's propagation speed\cite{kamata14}. The amplitude of dip A is about $0.01\%$ and does not decrease as $x$ increases. On the other hand, $\Delta t$ at which dip B appears behaves differently: the dip at $x=20$~$\mu$m is noticeably smaller than that close to the excitation gate at $x=5$~$\mu$m. Furthermore, $\Delta t$ at which the dip B appears is not proportional to $x$. These findings imply that in addition to the typical form of edge excitation known to occur in the Tomonaga-Luttinger liquid, there may also be alternative excitation mechanisms present close to the excitation gate. To discuss the details of this excitation, further experiments are required.

In summary, we reported on the dynamics of the FQH edge probed by stroboscopic PL and pump-probe measurements. While the time resolution of stroboscopic pump-probe measurement can be as short as the duration of the probe pulse ($\sim1$~ps), the time resolution of
stroboscopic PL measurement is constrained by the radiative recombination time of trions. Such an ultra-fast stroboscope measurement enables us to distinguish the normal mode of edge excitation from other higher energy non-linear excitations. The metric tensor $g_{\mu \nu}$ of the $(1+1)=2$-dimensional spacetime in the quantum universe simulator implemented in QH edges can be  obtained via a direct adaptation of this method. 

The authors are grateful to T. Fujisawa and M. Hotta for fruitful discussions. This work is supported by a Grant-in-Aid for Scientific Research (Grants Nos. 19H05603, 21F21016, 21J14386, and 21H05188) from the Ministry of Education, Culture, Sports, Science, and Technology (MEXT), Japan.

\textit{Data availability statement}: The data that support the
findings of this study are available upon reasonable request
from the author.

\bibliography{refs}

\end{document}